\documentclass[twocolumn,groupedaddress,
superscriptaddress,
showpacs,preprintnumbers,
amsmath,amssymb,
aps,
longbibliography
]{revtex4-1}

\usepackage{graphicx}
\usepackage{dcolumn}
\usepackage{bm}%
\usepackage[colorlinks=true,citecolor=blue]{hyperref}
\hypersetup{colorlinks=true,citecolor=blue,linkcolor=red,urlcolor=blue}

\usepackage{float}

\usepackage{color}
\definecolor{Green}{RGB}{0,204,102}
\definecolor{Purple}{RGB}{102,0,255}
\definecolor{Blue}{RGB}{51,153,255}
\definecolor{Red}{RGB}{151,010,010}

\begin{document}

\sloppy

\title{Tunable Low-Loss Hyperbolic Plasmon Polaritons in a T$_{d}\,$-WTe$_2$ Single Layer}

\author{Zahra Torbatian}
\affiliation{School of Nano Science, Institute for Research in Fundamental Sciences (IPM), Tehran 19395-5531, Iran}
\author{Dino Novko}
\email{dino.novko@gmail.com}
\affiliation{Institute of Physics, Bijeni\v{c}ka 46, 10000 Zagreb, Croatia}
\affiliation{Donostia International Physics Center (DIPC),
Paseo Manuel de Lardizabal 4, 20018 Donostia-San Sebasti\'an, Spain}
\author{Reza Asgari}
\email{asgari@ipm.ir}
\affiliation{School of Nano Science, Institute for Research in Fundamental Sciences (IPM), Tehran 19395-5531, Iran}
\affiliation{School of Physics, Institute for Research in Fundamental Sciences (IPM), Tehran 19395-5531, Iran}
\affiliation{ARC Centre of Excellence in Future Low-Energy Electronics Technologies, UNSW Node, Sydney 2052, Australia}

\begin{abstract}
Natural hyperbolic two-dimensional systems are a fascinating class of materials that could open alternative pathways to the manipulation of plasmon propagation and light-matter interactions. Here, we present a comprehensive study of the optical response in T$_d\,$-WTe$_2$ by means of density-functional and many-body perturbation theories. We show how monolayer WTe$_2$ with in-plane anisotropy sustains hyperbolic plasmon polaritons, which can be tuned via chemical doping and strain. The latter is able to extend the hyperbolic regime toward the near infrared with low losses. Moreover, with a moderate strain, WTe$_2$ can even be switched between elliptic and hyperbolic regimes. In addition, plasmons in WTe$_2$ are characterized by low losses owing to electron-phonon scattering, which is responsible for the temperature dependence of the plasmon line width. Interestingly, the temperature can also be utilized to tune the in-plane anisotropy of the WTe$_2$ optical response.
\end{abstract}

\maketitle

\section{Introduction}\label{sec:intro}
Optical properties of the material are considered hyperbolic when two of the principal components of the dielectric tensor are opposite in sign, i.e., one is metallic, with a negative dielectric constant, and the other dielectric, with regular transparent properties\,\cite{bib:smith03,bib:smith04,bib:krishnamoorthy12,bib:poddubny13,bib:gomezdiaz15,bib:gomezdiaz16, caldwell2014sub, dai2015graphene, brar2014hybrid}. This seemingly simple condition results in a plethora of exceptional optical properties, from low losses and large-wavevector response to the enhanced photonic density of states, which are not present in conventional elliptical materials. Most common forms of hyperbolic material present in the literature are artificially engineered, i.e., the so-called hyperbolic metamaterials (e.g., layered metal-dielectric structures)\,\cite{bib:poddubny13}.  However, the size of their components and a high degree of interface electron scatterings are limiting the corresponding imaging capabilities and resolution.

Hyperbolic materials with natural hyperbolic isofrequency surfaces in wavevector phase space were shown, on the other hand, to be characterized with low losses, high light confinement, and larger photonic density of states\,\cite{bib:sun14,bib:guan17,bib:ma18,bib:Li18,bib:zheng19}. The prerequisites for the appearance of natural hyperbolic surfaces are mainly anisotropic geometry and the peculiar interplay between intraband and interband electronic transitions. Under these circumstances the light-matter interaction could give birth to hyperbolic polaritons (i.e., plasmons, phonons, and excitons)\,\cite{bib:low16,bib:Li18,bib:ma18,bib:guo18,bib:zheng19,edalati2020mote2} with the aforesaid spectacular properties. Many layered anisotropic materials, such as graphite, MgB$_2$, cuprates, electrides, and transition metal dichalcogenides (TMDs)\,\cite{bib:sun14,bib:low16,bib:guan17,gjerding2017layered}, are expected to host these natural hyperbolic polaritons, while experimentally they were observed only in few of them, e.g., in MoO$_3$ surfaces\,\cite{bib:ma18,bib:zheng19} and structured hexagonal boron nitride\,\cite{bib:Li18}.

Two-dimensional (2D) crystalline materials supporting natural hyperbolic plasmon polaritons are considered even more attractive owing to their highly confined and tunable nature, e.g., with chemical doping, gating, or strain\,\cite{bib:low16,bib:wang20,bib:nemilentsau16,bib:lam15,bib:correasserrano16,bib:vanveen19}. These promising hyperbolic plasmonic surfaces were predicted to exist in 2D black phosphorous\,\cite{bib:correasserrano16,bib:vanveen19}, and very recently discovered in exfoliated T$_d\,$-WTe$_2$ thin films by means of Fourier transform-infrared spectroscopy\,\cite{wang2020van}. Moreover, experimental observations reveal low electron scatterings rates and decrease of anisotropy upon heating in both bulk and few-layers WTe$_2$\,\cite{homes2015optical,frenzel2017anisotropic,wang2020van}. The corresponding theoretical studies that could further corroborate and elucidate these findings are, however, still absent.
This semimetallic member of the TMDs family\,\cite{bib:kimura19} possesses also some other remarkable features, such as giant and anisotropic magnetoresistance\,\cite{ali2014large,pletikosic2014electronic,bib:na16,bib:wang16}, unusual transport properties\,\cite{soluyanov2015type,qian2014quantum,zheng2016quantum}, quantum spin Hall effect \cite{qian2014quantum,zheng2016quantum}, and superconductivity\,\cite{bib:sajadi18,bib:fatemi18}. All these renders WTe$_2$ an exciting new hyperbolic material of particular interest for planar nanophotonics and optoelectronics.

In this work, anisotropic 2D plasmon and hyperbolic plasmon dynamic of monolayer WTe$_2$ are studied in the framework of density functional theory (DFT) and many-body perturbation theory\,\cite{Novko2016,novko17,torbatian2020low}. Our analysis shows that the WTe$_2$ single layer is characterized with two (low- and high-energy) hyperbolic regions. The low-energy hyperbolic window is caused by the in-plane anisotropy in Drude weight (intraband excitations) and it was discussed in the experiment\,\cite{wang2020van}, while the high-energy region is due to highly anisotropic interband transitions at around 1\,eV.  We show how the hyperbolic properties in WTe$_2$ can be efficiently tuned by applying strain along the $a$-axis and doping. To this aim, we have considered doping of 0.1 electrons and 0.1 holes per unit cell as well as $\varepsilon=-4\%$ compressive and $\varepsilon=+4\%$ tensile strains. Such tuning induces modifications to the band structure of WTe$_2$ by making its electrodynamical properties more or less anisotropic. For instance, we show that for certain energies strain and doping can induce a crossover between hyperbolic and elliptic regimes. Furthermore, experimentally-observed low optical scattering rate and plasmon linewidth, as well as corresponding temperature dependence\,\cite{homes2015optical,wang2020van}, can be explained in terms of small electron-phonon coupling (EPC) present in WTe$_2$. The latter can be further modulated with strain and doping. Namely, the EPC strength could be increased by a factor of two when the compressive strain is applied. Finally, we demonstrate that anisotropy of plasmon dispersion and hyperbolicity in WTe$_2$ can be as well tuned with temperature. 

\section{THEORY AND COMPUTATIONAL METHODS}\label{sec:theo}
The {\it ab-initio} calculations are carried out in the framework of the local density approximation (LDA) of DFT within the QUANTUM ESPRESSO (QE) package \cite{0953-8984-21-39-395502}, using pz norm-conserving pseudopotentials with $6s^2 5d^4$ and $5s^2 5p^4$ valence electron configurations for W and Te, respectively.

\begin{figure*}[!t]
\begin{center}
\includegraphics[width=16.5cm]{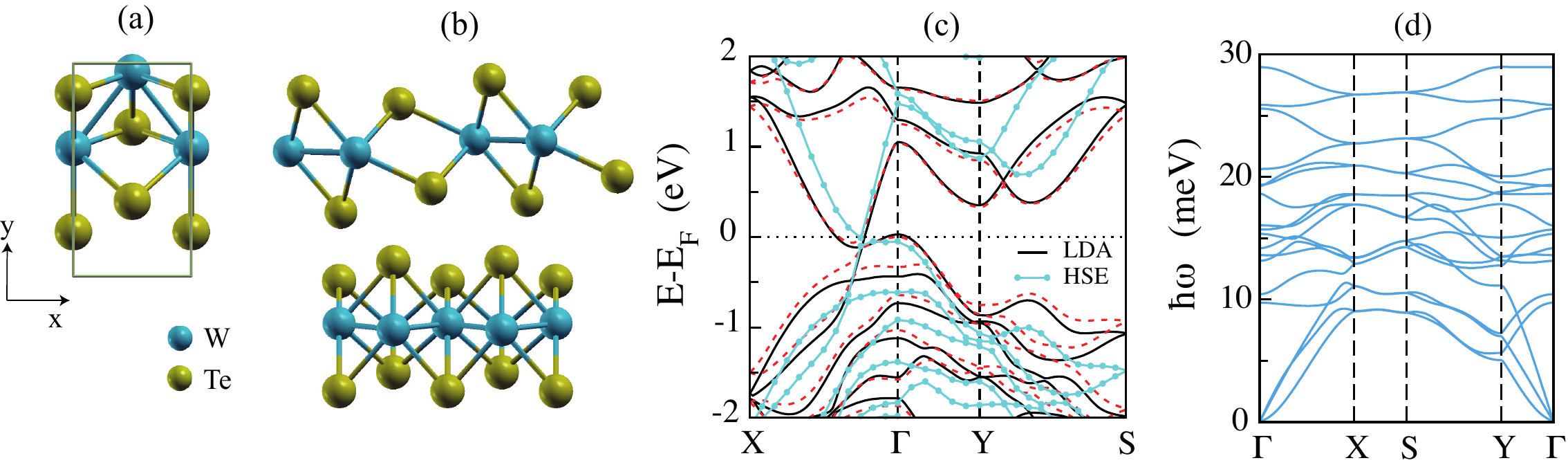}
\caption{(a) Top and (b) side views of monolayer T$_d$\,-WTe$_2$. W atoms are indicated with blue and Te atoms with green. The green rectangle marks the primitive unit cell. (c) The LDA electronic band structures with (dashed red lines) and without (solid black lines) spin-orbit coupling along high symmetry points. The band structure obtained with HSE hybrid functional without spin-orbit coupling is shown for comparison (blue dots). (d) Phonon spectrum of monolayer WTe$_2$. The Fermi energy (dotted line) is set to zero. }
\label{fig1}
\end{center}
\end{figure*}

We use a kinetic cutoff energy of $50.0$\,Ry, and a vacuum spacing of about $20\,{\rm \AA}$ is considered. The convergence criterion for energy is set to $10^{-5}$\,eV and the atomic positions are relaxed until the Hellmann-Feynman forces are less than $10^{-4}$\,${\rm eV/\AA}$. A set of $24 \times 12 \times 1$ ~$\Gamma$-centered $k$-point sampling is used for the primitive unit cell.
The optimized lattice constants are $a = 3.671\,{\rm \AA}$, $b = 6.642\,{\rm \AA}$, and both W and Te atoms occupy 2a Wyckoff positions corresponding to $(0,y,z)$ and $(0.5,-y, z+0.5)$ \cite{dawson1987electronic}.
The T$_d$  phase of monolayer WTe$_2$ [Figs.\,\ref{fig1}(a) and \ref{fig1}(b)]  is more stable than its H phase, not only on a substrate (in experiments) but also when it is freestanding. Therefore, we only concentrate on the T$_d$ phase of the compound.

\subsection{Theory of optical absorption and phonon-induced damping}\label{method}
We explore the optical absorption properties of WTe$_2$ by making use of the current-current response tensor calculated within DFT, where the electromagnetic interaction is mediated by the free-photon propagator. To do so, we pursue the same procedure given in Refs.\,\citenum{Novko2016,torbatian_PRB}.
First, we consider independent electrons which live in a local crystal potential obtained by DFT and interact with the electromagnetic field described by the vector potential. Then, we solve the Dyson equation for the screened current-current response tensor in the quasi-2D crystal of one or few layers
$\Pi=\Pi^0 + \Pi^0 \otimes D^0 \otimes \Pi$,
where $\Pi^0$ and $D^0$ are the non-interacting current-current response tensor and free-photon propagator, respectively.

The non-interacting current-current response tensor can be written as
\begin{eqnarray}
\Pi^0_{\mu}(\mathbf q, \omega)= \frac{2}{V} \sum_{\mathbf k, n, m} \frac{\hbar \omega}{E_n(\mathbf k)-E_m(\mathbf k+\mathbf q)}\nonumber\\
\times \left|J^\mu_{\mathbf k n , \mathbf k+\mathbf q m}\right|^2\frac{f_n(\mathbf k)-f_m(\mathbf k+\mathbf q)}{\hbar\omega + i\eta+ E_n(\mathbf k) - E_m(\mathbf k+\mathbf q)},
\label{current}
\end{eqnarray}
where $J^\mu_{\mathbf k n , \mathbf k+\mathbf q m}$ are the current vertices (see Refs.\,\citenum{Novko2016, torbatian_PRB} for more details) and $E_n(\mathbf k)$ are the Kohn-Sham energies.
Here $f_n(\mathbf k) $ is the Fermi-Dirac distribution at temperature $T$, and $V$ is the normalized volume. Further, the summation over $\mathbf k$ wavevectors is carried on a 120$\times$60$\times$1 grid, $n$ index sums over $30$ electronic bands, and  polarization directions are $\mu=x, y, z$.
Finally, the optical conductivity can be calculated as $\sigma_{\mu}(\omega)=-i\lim_{\mathbf{q}\rightarrow0}\Pi^0_{\mu}(\mathbf q, \omega)/\omega$, while the optical absorption is given by $A(\mathbf q,\omega)=-4\hbar {\rm Im}\,\Pi_{\mu}(\mathbf{q},\omega)/\omega$\,\cite{Novko2016,novko17,torbatian_PRB}.

To investigate the effects of phonons on the plasmon dispersion, we use the formalism presented in Refs.\,\citenum{novko17,Caruso2018}. Optical excitations are first convenient to decompose into the intraband ($n=m$) and interband ($n\ne m$) contributions. The electron-phonon scattering mechanism is then considered in the intraband channel.

For ${\mathbf q}\approx 0$, the intraband contribution of current-current response tensor can be written as the following~\cite{novko17,PhysRevB.3.305}:
\begin{eqnarray}
\Pi^0_{\mu}(\omega)= \frac{2}{V} \frac{\omega}{\omega[1+\lambda_{\rm ph}(\omega)]+i/\tau_{\rm ph}(\omega)}\sum_{\mathbf k, n} \frac{\partial f_{n\mathbf k}}{\partial E_{n\mathbf k}}  |J^\mu_{nn\mathbf k}|^2.
\label{intra_cu}
\end{eqnarray}
Here the effects of the EPC are contained in the temperature-dependent dynamical scattering time and energy renormalization parameters, i.e., $\tau_{\rm ph}(\omega)$ and $\lambda_{\rm ph}(\omega)$, respectively. The temperature-dependent dynamical scattering time is given by  \cite{novko2020broken,novko2018nonadiabatic}
\begin{eqnarray}
\hbar/\tau_{\rm ph}(\omega)=\frac{\pi\hbar}{\omega} \int d\Omega \alpha^2F(\Omega)\Big[2\omega \coth\frac{\Omega}{2k_BT}\nonumber\\-(\omega+\Omega)\coth\frac{\omega+\Omega}{2k_BT}+(\omega-\Omega)\coth\frac{\omega-\Omega}{2k_BT}\Big],
\label{tau}
\end{eqnarray}
where $k_B$ is the Boltzmann constant and
$\alpha^2F(\Omega)$ is the Eliashberg spectral function\,\cite{bib:giustino17,novko17,Caruso2018} and furthermore, the dynamical energy renormalization parameter $\lambda_{\rm ph}(\omega)$ is obtained by performing the Kramers-Kronig transformation of $1/\tau_{\rm ph}(\omega)$. The phonon properties (i.e., phonon energies and electron-phonon matrix elements), needed for calculating $\alpha^2F(\omega)$ and scattering rate Eq.\,\eqref{tau}, are obtained by means of density functional perturbation theory\,\cite{bib:baroni01} as implemented in QE. The $\alpha^2F(\omega)$ is calculated on 72$\times$36$\times$1 electron- and 24$\times$12$\times$1 phonon-momentum grids, respectively.

\begin{figure}[!t]
\begin{center}
\includegraphics[width=8.4cm]{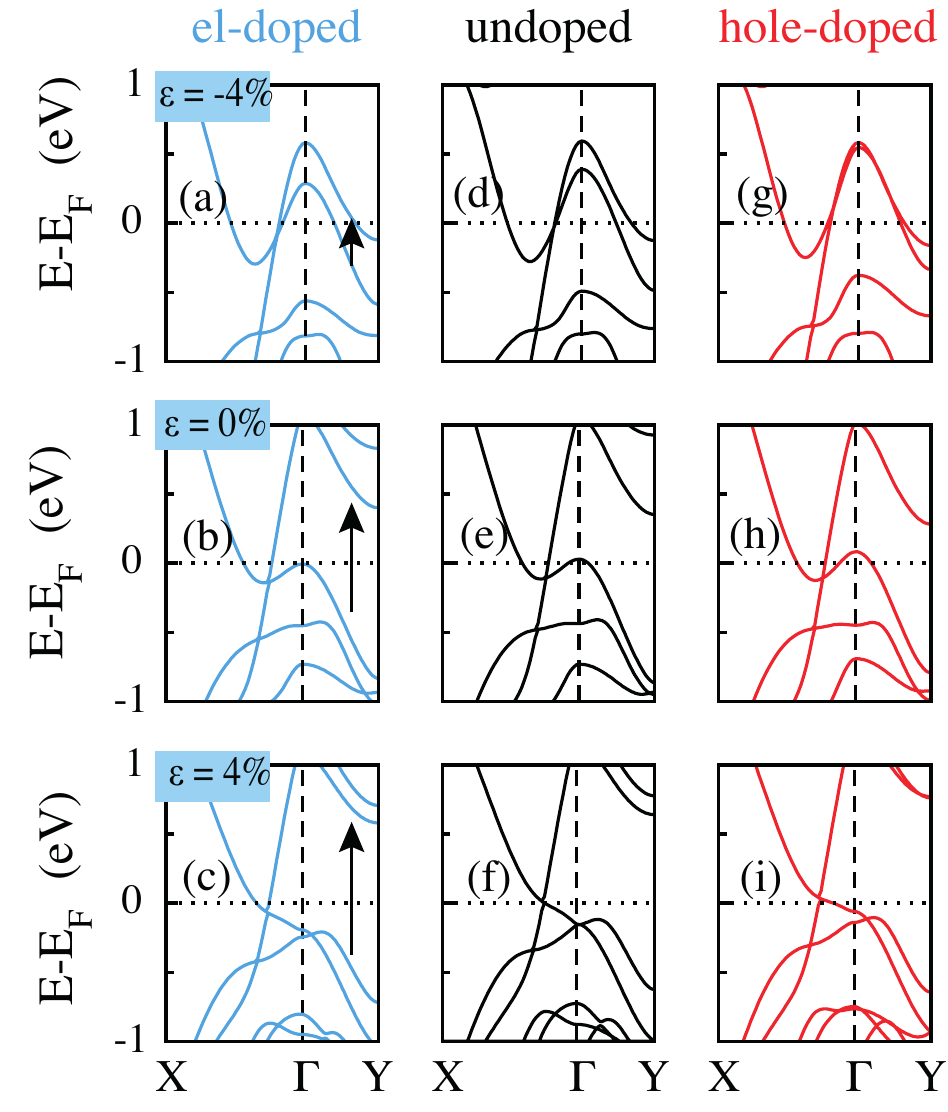}
\caption{The electronic band structures of monolayer WTe$_2$ along the high symmetry X-$\Gamma$-Y points for (a)-(c) 0.1 electron/u.c., (d)-(f) undoped and (g)-(i) 0.1 hole/u.c. and for $\varepsilon=-4\%$  (upper panels), 0$\%$  (middle   panels) and +4$\%$ (bottom panels) strains. Black arrows denote the interband transitions along the ${\rm \Gamma-Y}$ direction of the Brillouin zone. The Fermi energy is set to zero and indicated by dotted lines.}
\label{fig3}
\end{center}
\end{figure}

\section{RESULTS AND DISCUSSION}\label{sec:result}
\subsection{Electronic structure of monolayer WTe$_2$}

Unlike most TMDs which possess trigonal prismatic or monoclinic structures\,\cite{xu2013graphene}, WTe$_2$ adopts an orthorhombic type-II Weyl semimetallic (T$_d$) phase. The octahedron of Te atoms in monolayer WTe$_2$ is slightly distorted and the W atoms are displaced from their ideal octahedral sites, forming zigzag W-W chains along the $x$ direction as it is shown in Fig.\,\ref{fig1}(b). The distinct structural difference between $x$ and $y$ directions implies anisotropic in-plane properties.

\begin{figure*}[!t]
\begin{center}
\includegraphics[width=18cm]{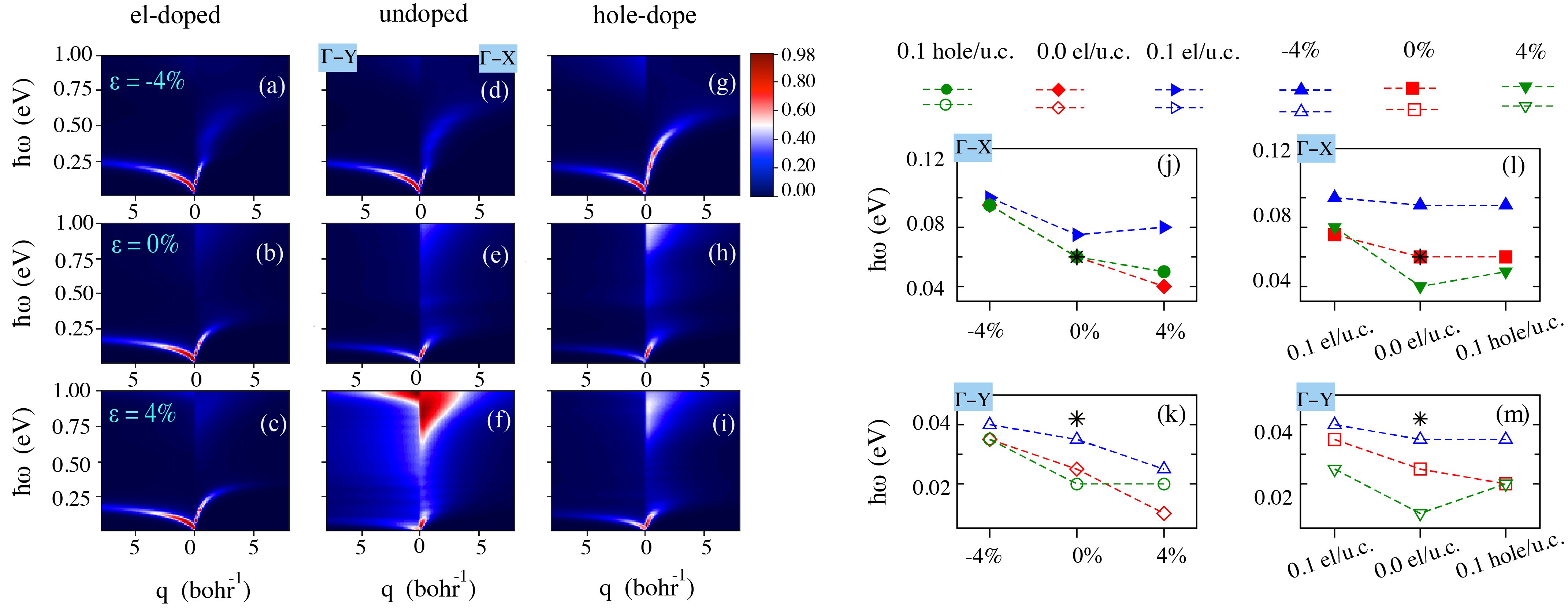}
\caption{The plasmon dispersions for (a)-(c) electron-doped (0.1 el/u.c.), (d)-(f) undoped, and (g)-(i) hole-doped (0.1 hole/u.c.) monolayer WTe$_2$ under $\varepsilon=-4\%$ compressive (upper panel), 0$\%$ (middle panel) and 4$\%$  tensile (bottom panel) strains when light is polarized along the $\Gamma-$X and $\Gamma-$Y directions. Notice, $q$ is in units of 10$^{-3}$ bohr$^{-1}$. (j)-(m) The plasmon peaks for the chosen small value of the wavevector $q$ ( $2\times10^{-4}\,$bohr$^{-1}$) along the $\Gamma-$X (upper panels) and $\Gamma-$Y (bottom panels) directions under different applied strains and for different dopings. The anisotropic plasmon energies in undoped WTe$_2$ is shown in (j)-(m) in comparison with the experiment\,\cite{wang2020van} (black stars).
}
\label{fig4}
\end{center}
\end{figure*}

The electronic band structure of monolayer WTe$_2$ along the high-symmetry points ${\rm X}-\Gamma-{\rm Y}-{\rm S}$ in the orthorhombic unit cell is shown in Fig.\,\ref{fig1}(c). The monolayer WTe$_2$ is a semimetal with the valence and conduction bands overlapped at the center of the hexagonal face in the Brillouin zone. In particular, there is a small electron pocket along $\Gamma-$X  and a small hole pocket around the $\Gamma$ point. This results in the characteristic tilted Weyl cone. Both the electron and hole pockets are corresponding to the zigzag W-W chain along the $a$ axis. The size of the electron pocket is almost the same as the hole pocket. The hole pocket originates from $5d$ orbital of W, while the electron pocket is formed by an avoided hybridization of a $5d$ band of W and $5p$ band of Te\,\cite{xiang2016quantum}.
Without spin-orbit coupling (SOC), the two bands cross at a point along the $\Gamma-$X line without any gap opening. However, in the presence of the SOC, the degeneracy at the crossing point is lifted, resulting in a small bandgap of approximately 50\,meV. Notice, we have not considered the SOC in our calculations for plasmon dispersion and for hyperbolic regimes owing to its small impact on optical spectra. However, these small changes are important in studying the temperature dependence of optical absorption (see below). 
For comparison, we also present the band structure obtained with Heyd–Scuseria–Ernzerhof (HSE) hybrid functional (blue dots). The main difference between the LDA and HSE band structures is in the value of the interband gap along $\Gamma-{\rm Y}$ direction, i.e., $\sim 1$\,eV for LDA and $\sim 1.5$eV for HSE. Our further analysis includes only the LDA band structure (since the HSE-based calculations of optical excitations would be computationally unfeasible). However, based on this band structure comparison, we do not expect important qualitative changes in our results if the HSE functional would be used instead.
Moreover, to examine the stability of undoped WTe$_2$, the phonon dispersion along the high-symmetry points in the Brillouin zone is calculated and shown in Fig.\,\ref{fig1}(d).

We tense and compress the relaxed unit cell along the $x$ direction in the strain range of $\varepsilon=4\%$ for undoped and doped (i.e., $0.1$ hole/u.c. and $0.1$ el/u.c. concentrations) WTe$_2$. Electron and hole dopings are simulated by adding and removing electrons and introducing the compensating homogeneous charged background.
The modulations in the band structure of monolayer WTe$_2$ by applying uniaxial strain along the lattice vector $a$ are displayed in Fig.\,\ref{fig3}. In the case of compressive strain, the monolayer WTe$_2$ keeps the semimetallic nature. The size of electron and hole pockets around the $\Gamma$  point (i.e., the Fermi surface) increases as the compressive strain is elevated. Under the tensile strain, the Fermi surface is significantly reduced and WTe$_2$ undergoes a phase transition from type-II to type-I semimetal. In other words, tensile strain changes the Weyl cone from tilted to normal. Besides the Fermi surface changes, the strain induces remarkable modifications of the interband threshold energy between the conduction and first valence band along the $\Gamma-{\rm Y}$ direction. Namely, the compressive strain can reduce the interband onset from $\sim 1$\,eV to below 100\,meV, while tensile strain is increasing it above 1\,eV [e.g., see black arrows in Figs.\,\ref{fig3}(a)-(c)]. Such modifications of Fermi surface and interband threshold energy in the $\Gamma-{\rm Y}$ direction alter the anisotropy of optical response in WTe$_2$.

It should be pointed out that the stability of monolayer WTe$_2$ under different strains and dopings is examined and checked.

\begin{figure*}[!t]
\begin{center}
\includegraphics[width=17.5cm]{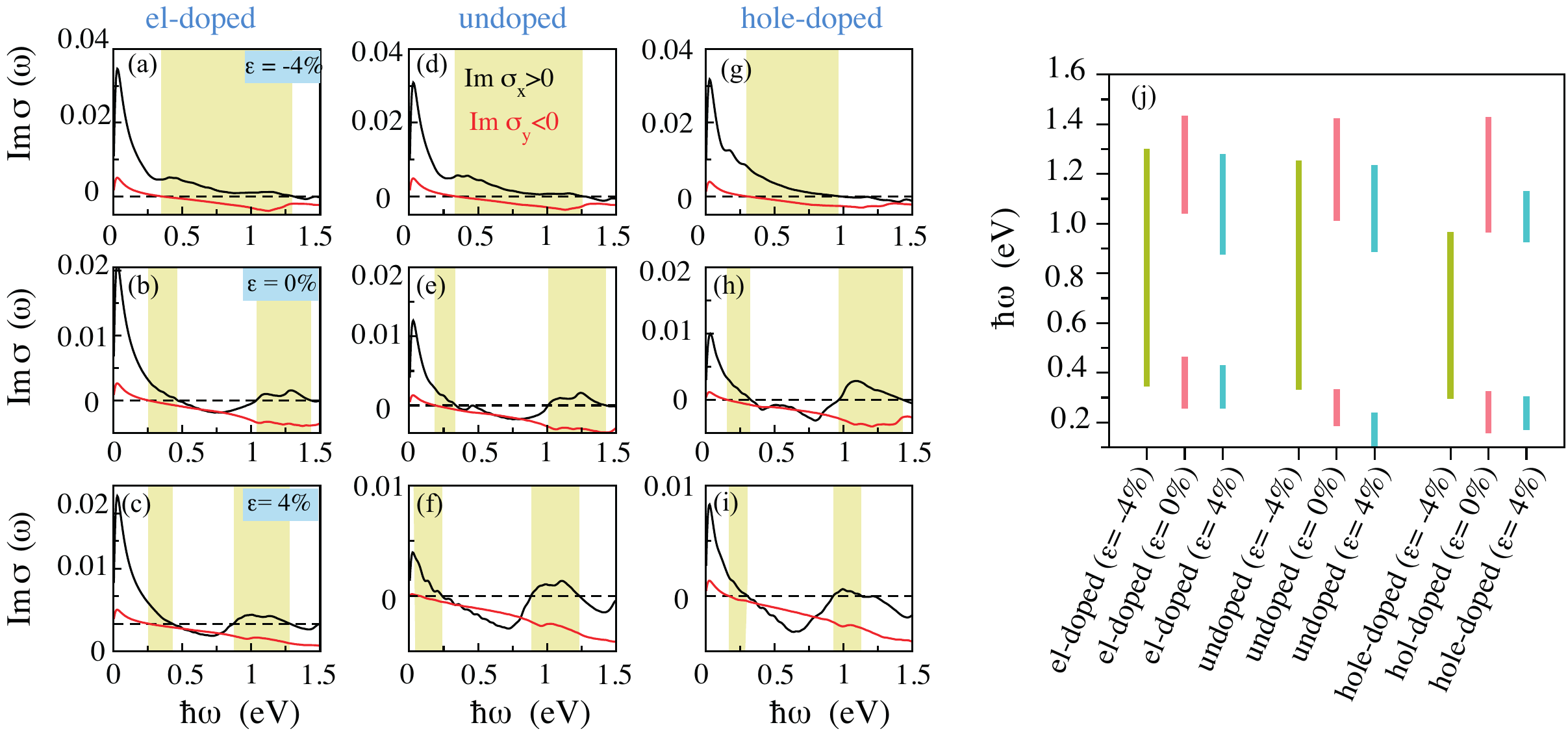}
\caption{Imaginary parts of optical conductivity along $x$ axis (black) and $y$ axis (red) for (a)-(c) 0.1 electron/u.c., (d)-(f) undoped, and (g)-(i) 0.1 hole/u.c., as well as under 4$\%$ compressive (upper panels), 0$\%$ (middle panels), and 4$\%$ tensile (bottom panels) strains. The hyperbolic regimes are depicted with shaded areas, where ${\rm Im}\,\sigma_{x}\times{\rm Im}\,\sigma_{y}<0$. (j) Calculated hyperbolic regimes of WTe$_2$ illustrated by wide lines. The hyperbolic regimes can be effectively tuned by changing the doping concentration and strain. For better comparison, the results under compressive, non- and tensile strains are illustrated with green, pink, and blue colors, respectively.
}
\label{fig6}
\end{center}
\end{figure*}

\subsection{Anisotropic plasmon dispersion}\label{sec:plasmon}

We study plasmon dispersion in undoped, electron- and hole-doped WTe$_2$ under strains of $\varepsilon=-4\%$ and $+4\%$ along the $a$ axis. The corresponding results are depicted in Fig.\,\ref{fig4}. In close agreement with the experiment\,\cite{wang2020van}, we obtain that light polarized along the $\Gamma-$X and $\Gamma-$Y directions induces two distinctive plasmon dispersions, which is the direct consequence of the in-plane anisotropy of the electronic structure in WTe$_2$. As also observed in the experiment\,\cite{wang2020van}, the intensity of the plasmon modes along the $a$ axis turns out to be larger than along the $b$ axis. Furthermore, the results clearly show that the plasmon modes can be tuned by means of strain and doping. By applying compressive strain the plasmon dispersion is blueshifted, while the tensile strain induces redshift. The latter behaviour can be directly related to the size of electron and hole pockets close to the $\Gamma$  point, which increases (decreases) with the compressive (tensile) strain. Indeed, the modifications of the band structure by strain manifest itself explicitly in the plasmon dispersion.

In order to show how strain and doping can effectively tune the plasmon dispersion in WTe$_2$, the plasmon peaks for the chosen, small value of wavevector $q$ ($2\times 10^{-4}$ bohr$^{-1}$) are plotted in Figs.\,\ref{fig4}(j)-(m). The results of the plasmon modes along the $\Gamma-$X ($\Gamma-$Y) direction as a function of strain and doping are shown in the upper (bottom) panels. Figures \ref{fig4}(a), \ref{fig4}(d) and \ref{fig4}(g) demonstrate how the plasmon dispersion of WTe$_2$ under compressive strain is less sensitive to doping compared to zero and tensile strains. On the other hand, plasmon dispersion under tensile strain shows the largest modifications due to doping. Furthermore, the undoped WTe$_2$ is most affected by strain, while the smallest effects are observed for the electron-doped case [Fig.\,\ref{fig4} (b)]. In addition, we emphasize that our results of anisotropic plasmon dispersion in undoped WTe$_2$ show a semi-quantitative agreement with the experiment\,\cite{wang2020van} [see black stars in Figs.\,\ref{fig4}(j)-(m)]. The slight deviation might come from the fact that we simulate plasmons for the freestanding system, while in the experiment the WTe$_2$ thin films are deposited on SiO$_2$/Si substrate, which might introduce additional doping.

\subsection{Hyperbolic regimes}\label{sec:hyperbolic}
We further demonstrate the existence of hyperbolic and normal (elliptical) regimes for certain spectral ranges. In terms of in-plane optical conductivity, the hyperbolic condition for specific photon energy $\hbar\omega$ is defined with
\begin{eqnarray}
\mathrm{Im}\,[\sigma_x(\omega)]\times \mathrm{Im}\,[\sigma_y(\omega)]<0.
\label{hypercond}
\end{eqnarray}
To this purpose, we discuss the possibility of tuning the range of hyperbolic region under strain and doping in WTe$_2$. Figure \ref{fig6} displays the imaginary parts of optical conductivities along two principal axes for different dopings and under $\varepsilon=\pm4\%$ compressive and tensile strains. The hyperbolic regimes, i.e., where the above condition is met, are depicted with the shaded area. The undoped WTe$_2$ single layer shows hyperbolic character in two separated regions, i.e., one starting at $\hbar \omega=0.18$\,eV and going up to $\hbar \omega=0.33$\,eV, and the other between 1.01\,eV and 1.42\,eV. The lower region comes from the anisotropy of Drude weight (intraband channel), while the higher region comes the interband transitions at $\sim 1$\,eV (which is only allowed for the light polarization along the $x$ direction).
In Ref.\,\citenum{wang2020van}, the lower hyperbolic regime is reported to be between 0.053\,eV and 0.078\,eV, while the higher regime was not investigated. The WTe$_2$ thin films as appearing in the experiment might have different (strained) unit cell parameters $a$ and $b$ as well as excess electron or hole charge due to the presence of the surface, and this might cause the differences between theory and experiment. Actually, in the following, we show how strain and doping can drastically modify the hyperbolic regime.

\begin{figure}[!t]
\begin{center}
\includegraphics[width=\columnwidth]{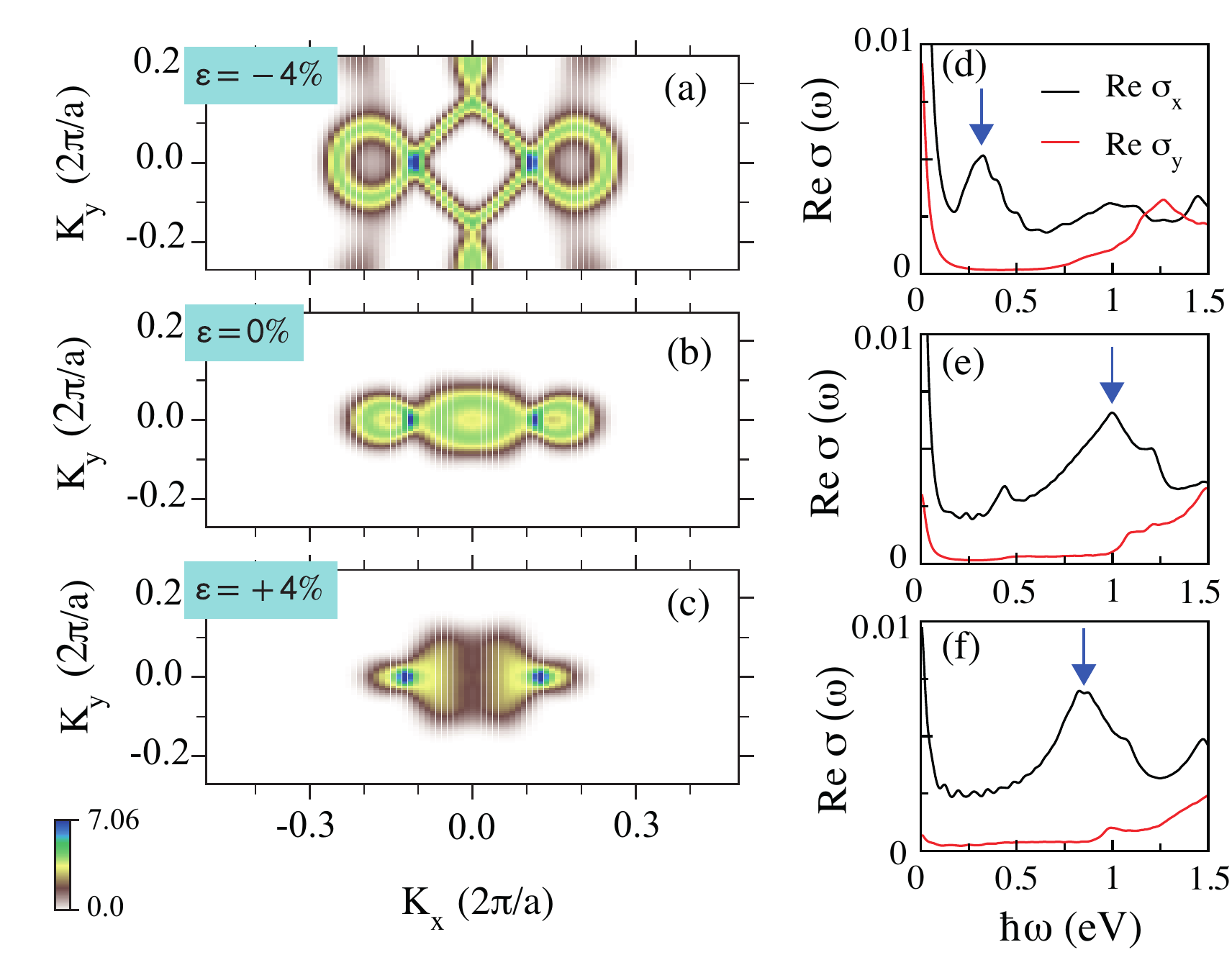}
\caption{Fermi surface plots for the undoped WTe$_2$ when (a) 4$\%$ compressive, (b) 0$\%$, and (c) 4$\%$ tensile strains are applied. Different colors represent different density of states at the Fermi level in units of ${\rm eV}^{-1}$. Real parts of optical conductivity for $x$ (black) and $y$ (red) polarizations of undoped WTe$_2$ and under (d) 4$\%$ compressive, (e) 0$\%$, and (f) 4$\%$ tensile strains. Blue arrows point to the peak of the interband transitions along the ${\rm \Gamma-Y}$ direction of the Brillouin zone, which are allowed only for the $x$ polarization and shift drastically with the strain.
}
\label{fig7}
\end{center}
\end{figure}

By applying compressive strain along the $x$ direction (upper panels of Fig.\,\ref{fig6}) the energy windows of the two hyperbolic regimes are significantly extended, where the large elliptical region ranging from around 0.4\,eV to 1\,eV is modified to hyperbolic.
Therefore, with a moderate strain, WTe$_2$ can even be switched between elliptic and hyperbolic materials. Under the tensile strain, the two hyperbolic regions present for the undoped case are modified, however, they are not joined into the single large energy window as it is the case for the compressive strain. As mentioned earlier, these dramatic changes in hyperbolic regions are due to the Fermi surface and interband threshold modifications.

\begin{figure*}[!t]
\begin{center}
\includegraphics[width=15cm]{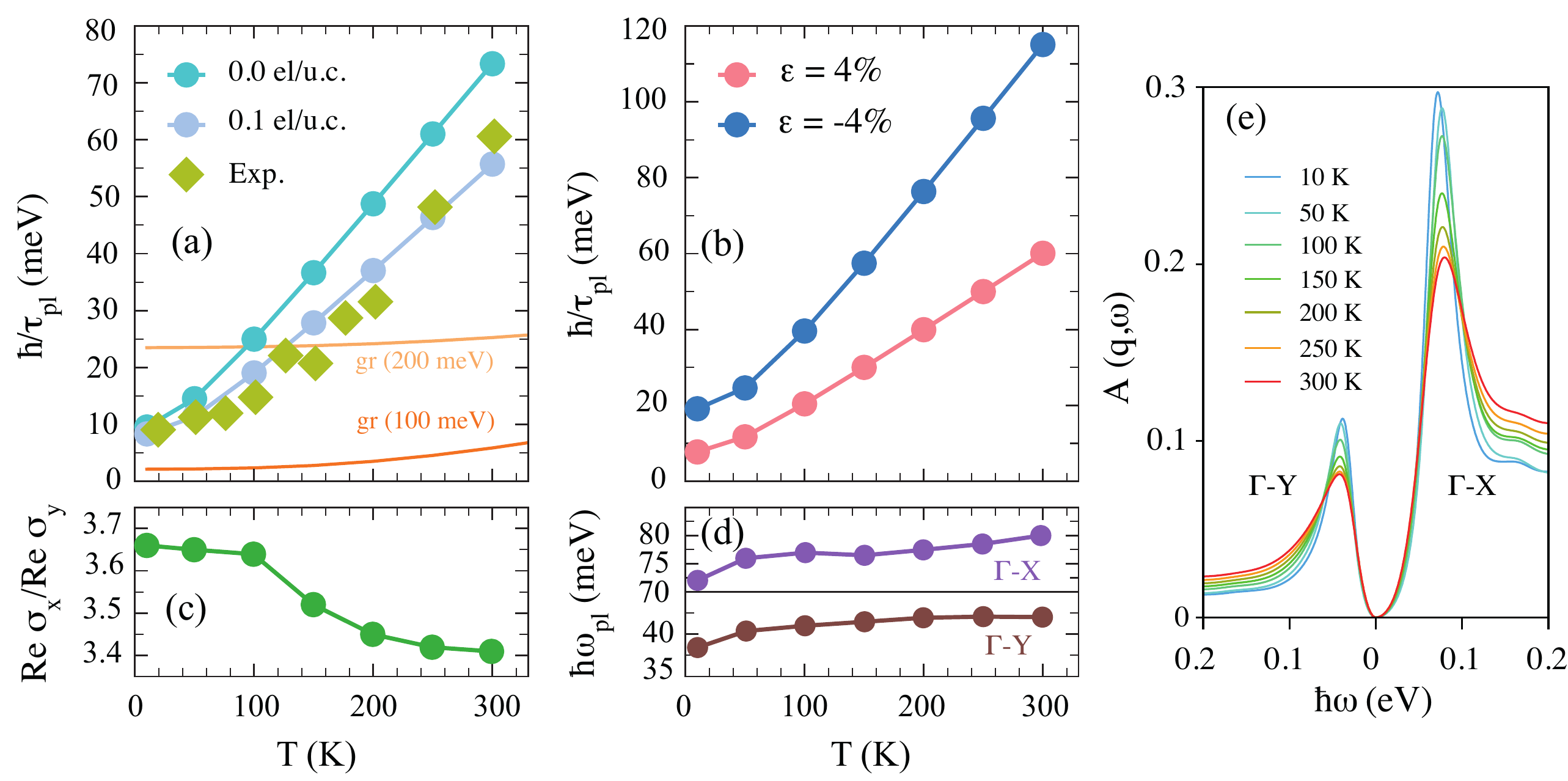}
\caption{(a) Calculated temperature dependence of plasmon linewidth due to electron-phonon coupling for pristine and electron-doped WTe$_2$ single layer. Experimental data are taken from Ref.\citenum{wang2020van}. For comparison, plasmon linewidths due to electron-phonon scatterings are shown for electron-doped graphene (0.1\,el/u.c.) when plasmon energy is 100\,meV and larger than 200\,meV. (b) Plasmon linewidth as a function of temperature for electron-doped WTe$_2$ under $+4\%$ and $-4\%$ strains. (c) The ratio between static conductivities along the $x$ and $y$ directions as a function of temperature. (d) The plasmon energy peaks along the both in-plane principal directions as a function of temperature ($q=2\times10^{-4}$ bohr$^{-1}$).
(e) Temperature dependence of optical absorption $A(q,\omega)$ for $q=2\times10^{-4}$ bohr$^{-1}$ of undoped WTe$_2$ along the ${\rm \Gamma-X}$ and ${\rm \Gamma-Y}$ directions.}
\label{fig8}
\end{center}
\end{figure*}

In Fig.\,\ref{fig6}(j) we explicitly show the extent of the hyperbolic energy windows for different strains and dopings. It is evident that WTe$_2$ single layer is hyperbolic over a broad energy window from far- to near-infrared (0.08\,eV to 1.43\,eV). Moreover, we observe that under the tensile strain, the hyperbolic region in undoped, electron- and hole-doped WTe$_2$ is redshifted, especially the second high-energy hyperbolic window (energies lager than 1.0\,eV). Indeed, under strain, optical absorption peaks shift to lower energies [Fig.\,\ref{fig4}(c), \ref{fig4}(f), and \ref{fig4}(i)],  and as a result, the onset of the hyperbolic region goes down. Furthermore, the onset of the hyperbolic region can be tuned by changing the doping concentration, i.e., it is redshifted (blueshifted) for hole-doped (electron-doped) case.

In Fig.\,\ref{fig7} we elaborate a bit more on the origin of these hyperbolic regions and their sensitivity on compressive and tensile strains. Figures \ref{fig7}(a)-(c) show the Fermi surface modifications of undoped WTe$_2$ with respect to strain. It is clear from these results that the Fermi surface is drastically increased for $\epsilon=-4\,\%$, while decreased alomst to a single point for $\epsilon=+4\,\%$. By applying strain, the anisotropy of Fermi surface is changed, however it remains for all the cases considered here. The Fermi surface modifications can also be seen in terms of different Drude tails of ${\rm Re}\,\sigma_x(\omega)$ and ${\rm Re}\,\sigma_y(\omega)$, which are shown in Figs.\,\ref{fig7}(d)-(f). The anisotropy of these Drude tails results in the low-energy hyperbolic region. Optical absorption spectra for $x$ polarization ${\rm Re}\,\sigma_x(\omega)$ of undoped WTe$_2$ are characterized with the interband peak around 1\,eV, which does not appear for the $y$ polarization. As discussed before, the origin of this prominent peak are interband transitions along ${\rm \Gamma-Y}$ path in the Brillouin zone (see black arrows in Fig.\,\ref{fig3}). The results show that these interband peak is highly sensitive on the applied strain. From this it is evident that the second, high-energy hyperbolic region is caused by these interband transitions, which are optically active for $x$, while inactive for the $y$ polarizations. Note also that the onset of the high-energy hyperbolic region coincide with the peak position of this interband transition in each of the presented cases.

We note that the hyperbolic surface in the WTe$_2$ single layer turns out to be even more sensitive to compressive strain than in the black phosphorus\,\cite{bib:vanveen19}.


\subsection{Electron-phonon coupling and temperature dependence of plasmon mode}\label{sec:temp}
For completeness, we investigate the effect of the EPC on hyperbolic plasmon polaritons in WTe$_2$ and the corresponding temperature effects. The strain- and doping-induced modifications of electron-phonon interaction are also discussed. 

The EPC constant in undoped unstrained WTe$_2$ single layer turns out to be $\lambda=0.43$, and as a result the phonon-induced plasmon decay rate (plasmon linewidth) at $T=10$\,K and for plasmon energies $\hbar\omega_{\rm pl}> 100$\,meV is $\hbar/\tau_{\rm pl}=9.5\,$meV. These values for $\lambda$ and $\hbar/\tau_{\rm pl}$ are almost unaltered when tensile strain or electron doping are applied. Interestingly, when the compressive strain is applied the EPC strength and plasmon linewidth are enhanced to $\lambda=0.70$ and $\hbar/\tau_{\rm pl}=19\,$meV, respectively. This is due to significant Fermi surface increase, as it can be seen in Figs.\,\ref{fig3}(a), \ref{fig3}(d), and \ref{fig3}(g).


Furthermore, Fig.\,\ref{fig8}(a) shows temperature dependence of plasmon linewidth due to EPC of undoped and electron-doped WTe$_2$. These results show an excellent agreement with the Drude scattering rate extracted from the experiment\,\cite{wang2020van}, suggesting that the dominant contribution to the Drude scattering as well as plasmon damping rates is due to coupling with phonons. Very similar temperature dependence was obtained for bulk WTe$_2$\,\cite{homes2015optical}, where the thermal enhancement of the scattering rate was attributed to electron scatterings within Fermi liquid (i.e., electron-electron scattering). Here we show, however, that the EPC might be more important than electron-electron scatterings in WTe$_2$.

For comparison, we also plot in Fig.\,\ref{fig8}(a) the plasmon linewidth as a function of temperature for a prototypical plasmonic material, namely, electron-doped graphene, when plasmon energy is $\hbar\omega_{\rm pl}=100$\,meV (orange) and $\hbar\omega_{\rm pl}> 200$\,meV (light orange)\,\cite{bib:novko20}. Compared to graphene, the plasmon linewidth in WTe$_2$ shows a much steeper increase with temperature, which is because the phonons in WTe$_2$ have energies up to $\sim 30$\,meV [see Fig.\,\ref{fig1}(d)] and are thus energetically available already around room temperature. On the other hand, in graphene only the weakly coupled acoustic phonons are thermally excitable at low temperatures, while the strongly coupled optical phonons have energies between 160\,meV, and 200\,meV and thus cannot contribute\,\cite{novko17,bib:novko20}. For energies of $\sim 100$\,meV the plasmons in graphene have a much smaller decay rate compared to WTe$_2$. However, for larger excitation energies (i.e., 200\,meV or more) when the strongly coupled phonons are activated in graphene, the WTe$_2$ plasmons are less damped for certain temperatures. Moreover, Fig.\,\ref{fig8}(b) displays how strain can modify significantly the plasmon linewidth and its temperature dependence for 0.1\,el/u.c. For instance, at around room temperature, the plasmon linewidth is increased from 60\,meV to 115\,meV by changing strain from $\varepsilon=-4\%$ to $+4\%$. As discussed earlier, this increase is due to modifications of the Fermi surface, which then changes the EPC strength.

Finally, we demonstrate how temperature can drastically modify the plasmon energies in undoped WTe$_2$, and thus decrease the anisotropy and hyperbolic properties. The plasmon energy of 2D system along $x$ or $y$ directions can be calculated as\,\cite{bib:kupcic14}:
\begin{eqnarray}
\omega_{\rm pl}^2(\mathbf{q},\omega) = 2\pi q_{\mu} \omega {\rm Im}\left[ \sigma_{\mu}^{\rm intra}(\mathbf{q},\omega)+\sigma_{\mu}^{\rm inter}(\mathbf{q},\omega) \right].
\label{eq:plen}
\end{eqnarray}
The temperature dependence of the plasmon energy comes from the Fermi-Dirac distribution function that enters the intraband and interband conductivities, i.e., $\sigma_{\mu}^{\rm intra}(\mathbf{q},\omega)$ and $\sigma_{\mu}^{\rm inter}(\mathbf{q},\omega)$, but also from the EPC via the $1/\tau_{\rm ph}(\omega)$ and $\lambda_{\rm ph}(\omega)$ functions.
First, in Fig.\,\ref{fig8}(c) we show the ratio between static ($\omega=0$) conductivities along the $x$ and $y$ directions, i.e.,  ${\rm Re}\,\sigma_{x} / {\rm Re}\,\sigma_{y}$, as a function of temperature. Since one can also write ${\rm Re}\,\sigma_{x} / {\rm Re}\,\sigma_{y} = m^{\rm eff}_y / m^{\rm eff}_x$, this also measures the anisotropy of the effective masses along the two directions. The results demonstrate how the anisotropy can be reduced with moderate heating. This is in agreement with Refs.\,\citenum{frenzel2017anisotropic,wang2020van}. However, in these experiments the overall anisotropy of $m^{\rm eff}$ at $T=10$\,K is smaller, and the temperature-induced reduction of anisotropy is greater. This discrepancy might come from the fact that we simulate single layer WTe$_2$, while in these experimental studies the authors investigate few-layer and bulk WTe$_2$ samples, where additional interband transitions might be present. We note that we have introduced the SOC here in order to slightly improve the agreement with the experiments.
In Figs.\,\ref{fig8}(d) and \ref{fig8}(e) we plot, respectively, the plasmon energy peaks and optical absorption spectra for $q=2\times10^{-4}$ bohr$^{-1}$ along the $\Gamma-$X and $\Gamma-$Y directions as a function of temperature. When the temperature increases the plasmon intensities along the both axes are reduced, however, the intensity reduction along the $x$ axis is more considerable. This is again in agreement with the experimental observations\,\cite{wang2020van}. The energies of plasmons along the two directions are also modified with increasing temperature.

All in all, as temperature increases, the anisotropy of plasmonic features along both in-plane axes is reduced. Our theoretical analysis shows that this comes from the temperature-induced modifications of the EPC as well as of the electron distribution. This shows how the hyperbolic features in WTe$_2$ can be tuned even with temperature, which makes it a highly attractive and tunable hyperbolic material.

\section{CONCLUSION}\label{sec:concl}
By applying first-principles theory we have demonstrated that monolayer T$_d$\,-WTe$_2$ is a natural type hyperbolic material with low losses, where the hyperbolic regime can be present from far- to near-infrared frequencies. Moreover, the chemical doping and strain offer convenient methods to effectively tune the hyperbolicity in WTe$_2$. For instance, with a moderate strain, WTe$_2$ can even be switched between elliptic and hyperbolic regimes at certain frequencies.
Remarkable modifications of optical response are a direct consequence of strain- and doping-induced peculiar alterations of the Fermi surface and interband threshold energy.
We have also calculated the electron-phonon coupling in WTe$_2$, which can also be tuned with doping and strain. Our analysis shows that the temperature increase of the plasmon linewidth and Drude scattering rate observed in the experiments can be successfully explained in terms of electron-phonon interaction. The latter is actually relatively small in WTe$_2$ and, therefore, the phonon-induced plasmon decay rate is smaller for certain energies and temperatures than the decay rate of the plasmon in prototypical graphene. Finally, we have shown that a slight temperature increase can alter the anisotropy of optical response along both in-plane axes and change thus the hyperbolic regime.

The presented methodology is also expected to help elucidate hyperbolic optical response in similar novel materials with strong anistropic in-plane electronic properties, such as T$_d$\,-MoTe$_2$\,\cite{bib:lai18}, TaIrTe4\,\cite{bib:lai18b}, or black phosphorus analog SnSe\,\cite{bib:pretikosic18}.

\begin{acknowledgments}
This work is supported by the Iran Science Elites Federation.
D.N. acknowledges financial support from the Croatian Science Foundation (Grant no. UIP-2019-04-6869) and from the European Regional Development Fund for the ``Center of Excellence for Advanced Materials and Sensing Devices'' (Grant No. KK.01.1.1.01.0001).
\end{acknowledgments}

\bibliography{wte2}

\end{document}